\documentclass[twocolumn, longbilbiography, showkeys]{aastex63}

\usepackage{rotating}
\usepackage{acronym}
\usepackage{xspace}
\usepackage{multirow}
\usepackage{mathtools}
\usepackage{amsmath}
\usepackage{hyperref}
\usepackage{units}
\usepackage{ulem}
\usepackage{physics}

\usepackage{xcolor}

\graphicspath{{./}{figures/}}

\def\chieff{\ensuremath{\chi_\mathrm{eff}}\xspace}
\def\q{\ensuremath{q}\xspace}
\def\z{\ensuremath{z}\xspace}

\acrodef{GW}{gravitational wave}
\acrodef{BH}{black hole}
\acrodef{BBH}{binary black hole}
\acrodef{NS}{neutron star}
\acrodef{BNS}{binary neutron star}
\acrodef{EM}{electromagnetic}

\acrodef{LIGO}{Laser Interferometer Gravitational-wave Observatory}
\acrodef{LVC}{LIGO Scientific and Virgo Collaboration}
\acrodef{LVK}{LIGO--Virgo--KAGRA Collaboration}
\acrodef{O1}{first observing run}
\acrodef{O2}{second observing run}
\acrodef{O3}{third observing run}
\acrodef{O3a}{first half of the third observing run}

\newcommand{\CIT}{\affiliation{Department of Physics, California Institute of Technology, Pasadena, CA 91125, USA}}
\newcommand{\CITLab}{\affiliation{LIGO Laboratory, California Institute of Technology, Pasadena, CA 91125, USA}}
\newcommand{\TAPIR}{\affiliation{TAPIR, California Institute of Technology, Pasadena, CA 91125, USA}}
\newcommand{\Adler}{\affiliation{Adler Planetarium, 1300 South DuSable Lake Shore Drive, Chicago, IL, 60605, USA}}
\newcommand{\CIERA}{\affiliation{Center for Interdisciplinary Exploration and Research in Astrophysics (CIERA), Northwestern University, Evanston, IL, 60201, USA}}

\renewcommand{\chieff}{\chi_\textrm{eff}}
\newcommand{\chip}{\chi_{p}}
\newcommand{\chit}{\chi_\textrm{thres}}

\shorttitle{How to diagnose a hierarchical merger origin}
\shortauthors{Payne, Kremer, and Zevin}

\begin{document}

\title{Spin Doctors: How to diagnose a hierarchical merger origin}

\author[0000-0003-4507-8373]{Ethan Payne}
\email{epayne@caltech.edu}
\CIT
\CITLab

\author[0000-0002-4086-3180]{Kyle Kremer}
\thanks{NASA Hubble Fellow}
\TAPIR

\author[0000-0002-0147-0835]{Michael Zevin}
\Adler
\CIERA

\begin{abstract}
Gravitational-wave observations provide the unique opportunity of studying black hole formation channels and histories --- but only if we can identify their origin. 
One such formation mechanism is the dynamical synthesis of black hole binaries in dense stellar systems. 
Given the expected isotropic distribution of component spins of binary black hole in gas-free dynamical environments, the presence of anti-aligned or in-plane spins with respect to the orbital angular momentum is considered a tell-tale sign of a merger's dynamical origin. 
Even in the scenario where birth spins of black holes are low, hierarchical mergers attain large component spins due to the orbital angular momentum of the prior merger. 
However, measuring such spin configurations is difficult. 
Here, we quantify the efficacy of the spin parameters encoding aligned-spin ($\chieff$) and in-plane spin ($\chip$) at classifying such hierarchical systems.  
Using Monte Carlo cluster simulations to generate a realistic distribution of hierarchical merger parameters from globular clusters, we can infer mergers' $\chieff$ and $\chip$.
The cluster populations are simulated using Advanced LIGO-Virgo sensitivity during the detector network's third observing period and projections for design sensitivity. 
Using a ``likelihood-ratio''-based statistic, we find that $\sim2\%$ of the recovered population by the current gravitational-wave detector network has a statistically significant $\chip$ measurement, whereas no $\chieff$ measurement was capable of confidently determining a system to be anti-aligned with the orbital angular momentum at current detector sensitivities. 
These results indicate that measuring spin-precession through $\chip$ is a more detectable signature of a hierarchical mergers and dynamical formation than anti-aligned spins.
\end{abstract}


\section{Introduction}\label{sec:intro}
Following the first handful of observations of \ac{BBH} mergers through their \ac{GW} emission~\citep{GW150914, GWTC1, GWTC2}, many studies predicted that the dominant formation channel of \acp{BBH} would be determined after $\mathcal{O}(10-100)$ observations~\citep{Zevin2017,Stevenson2015,Stevenson2017,Fishbach2017WhereAreBigBH,Vitale2017,Gerosa2017,ArcaSedda2019,Safarzadeh2020, Gerosa2021}.
However, despite the \ac{LVK} detector network accumulating nearly $100$ confident \ac{BBH} observations~\citep{GWTC3}, prominent formation pathways for \ac{BBH} mergers remains an open question in \ac{GW} astrophysics. 
The incongruity between prior expectation and reality can be attributed to a number of factors: 
\begin{enumerate}
    \item The diversity in the gravitational-wave events detected thus far does not show a strong preference for any one formation channel, with observations spanning a broad range of masses and mass ratios~\citep[e.g.][]{GWTC1,GWTC2,GWTC3, Olsen2022pin, Mehta2023zlk}. 
    \item Additional potential formation channels have been proposed in addition to the canonical ``dynamical-versus-isolated'' distinction~\citep[see e.g.][for a review]{Mandel2022}, as well as subchannels to these canonical birth environments, which muddles the ability to pin down specific birth environments~\citep{Cheng2023}. 
    \item Uncertainties in massive-star evolution, binary physics, and formation environments are more vast than previously appreciated, translating to larger uncertainties in expected parameter distributions and generally making inference difficult~\citep[see e.g.][for reviews]{Mapelli2021,Spera2022}. 
    \item Unlike \acp{BH} in high-mass X-ray binaries in the Milky Way, which have been argued to have spin estimates that are near extremal~\citep{Liu2008,MillerJones2021,Reynolds2021}, the population of spins for \ac{GW}-detected \acp{BH} are relatively small~\citep{LVKpopsO3}, making it difficult to distinguish between small, aligned spins expected from isolated evolution and moderate, in-plane spins expected from dynamical assembly.
\end{enumerate}
In addition to spins, trends in the mass spectrum~\citep[e.g.][]{Stevenson2015,Zevin2017, Fishbach2021, GWTC3pop,Belczynski2022,Mahapatra2022,VanSon2023}, redshift evolution~\citep[e.g.][]{RodriguezLoeb2018,vanSon2022,Fishbach2023}, orbital eccentricity~\citep[e.g.][]{Zevin2021Eccentricity}, and correlations between \ac{BBH} parameters~\citep[e.g.][]{Callister2021,Tiwari2021,Zevin2022SuspiciousSiblings,Broekgaarden2022,McKernan2022, Adamcewicz2022,Baibhav2022, Biscoveanu2022, Ray2023} have been investigated to elucidate the contribution of the various proposed \ac{BBH} formation channels, although a robust conclusion is still far from being reached. 

Although the holistic approach of examining features of the \textit{full} \ac{BBH} population holds promise for constraining formation scenarios~\citep{Zevin2021OneChannel}, a complementary approach is the identification of \textit{individual} merger events with distinguishing features uniquely associated with one or a subset of formation pathways. 
One example of this is eccentricity: \ac{BBH} mergers with measurable eccentricity 
in the \ac{LVK} sensitive frequency range ($\gtrsim 0.05$ at 10 Hz, \citealt{Lower2018,RomeroShaw2019}) strongly point to a recent dynamical interaction, as orbital eccentricity quickly dissipates if a \ac{BBH} system inspirals over a long timescale. 
Although no eccentric \ac{BBH} mergers have been confidently detected to date (though see \citealt{RomeroShaw2022}), the detection of a small number of eccentric mergers (or non-detection of eccentric mergers) would place stringent constraints on the contribution of dynamical formation pathways~\citep{Zevin2021Eccentricity}. 

Another possible smoking-gun signal of dynamical formation is the presence of hierarchical mergers --- \ac{BBH} mergers where one or both of the component \acp{BH} have gone through a previous merger event. 
Hierarchical mergers have masses that are typically larger than their ``first-generation'' progenitors that were born from massive stars 
as well as distinctive signatures in their spin magnitudes ($a \approx 0.7$, with a dispersion based on the mass ratio and component spins of the prior merger) and spin orientations (an isotropic distribution assuming a gas-free dynamical formation environment). 
Although hierarchical mergers are predicted to contain black holes with masses in the (pulsational) pair instability mass gap and studies have attempted to quantify the likelihood of particular \ac{GW} systems being hierarchical merger~\citep{Kimball2019,Kimball2020,Kimball2021, Mahapatra2021}, uncertainties in the size and location of the gap~\citep{Farmer2019, GWTC3pop, Edelman2021}, measurement uncertainties for high-mass black holes~\citep{GW190521}, and prior considerations~\citep{Fishbach2020,Nitz2021, Mould2023} make mass alone difficult to pin down whether a particular system contains a black hole that was the result of a prior merger. 

To identify the tell-tale signatures of hierarchical mergers, it is useful to consider the leading-order (i.e., typically best-measured) spin terms from the post-Newtonian expansion of the \ac{GW} waveform: the \textit{effective spin}~\citep{Damour2001, Racine2008}
\begin{equation}\label{eq:chieff}
    \chieff = \frac{a_1\cos\theta_1 + qa_2\cos\theta_2}{1+q},
\end{equation}
and \textit{precessing spin}~\citep{Schmidt2015}
\begin{equation}\label{eq:chip}
    \chip = \textrm{max}\Big(a_1\sin\theta_1, q\frac{3+4q}{4+3q}a_2\sin\theta_2 \Big),
\end{equation}
parameters where $q$ is the mass ratio between the secondary and primary black holes, and $a_1$ and $a_2$ are the primary and secondary black holes' spins, respectively. 
The effective spin encodes a mass-weighted projection of the spin vectors on the orbital angular momentum axis, whereas $\chip$ depends on the projection of the spin vector on the plane of the orbit and is related to the strength of precession of the orbital angular momentum about the total angular momentum. 

Hierarchical mergers are expected to have distinctive signatures in both of these spin parameters; due to generally large spin magnitudes (acquired during their first generation merger, \citealt{Buonanno2008,Fishbach2017Hierarchical}) and isotropic spin orientations (a natural feature of dynamical formation in gas-poor environments, e.g. \citealt{Rodriguez2019}), some hierarchical mergers should show evidence for negative $\chieff$, and others for large $\chip$. 
While a positive $\chieff$ is possible, such systems may not be distinguishable from other formation channels whereas spin anti-alignment is difficult to form in the field~\citep{Rodriguez2016}. 
Being a typically better-measured parameter~\citep{Ng2018, Biscoveanu2021}, studies have focused on negative $\chieff$ as a potential sign for a hierarchical merger event \citep[e.g.,][]{Baibhav2020, Fishbach2022, Zhang2023}. 
However, due to the inherent isotropic spin orientation distribution that is expected for hierarchical mergers in most astrophysical environments, many more systems will have large in-plane spins as opposed to large spins anti-aligned with the orbital angular momentum. 
For example, from cluster population simulations (see Sec.~\ref{sec:cmc}), $\sim$\,$0.5\%$ ($\sim$\,$20\%$) of hierarchical systems will have $\chieff<-0.5$ ($\chieff < -0.2$) whereas $\sim$\,$67\%$ ($\sim$\,$96\%$) of systems will have $\chip>0.5$ ($\chip > 0.2$). 
So while $\chieff$ is expected to be better measured, a significantly higher fraction of the hierarchical population will have the distinct signature of precession. 

In this letter, we investigate the ability to measure each of these parameters for the purpose of identifying specific \ac{BBH} mergers that are likely of a hierarchical origin. 
We take synthetic \ac{BBH} mergers from realistic models of globular clusters, performing full parameter estimation on $6\times10^3$ events. 
Using these realistic measurement uncertainties, we quantify the fraction of hierarchical mergers that confidently exhibit negative $\chieff$ and large $\chip$. 
Despite larger typical measurement uncertainties, we show that $\chip$ is a better indicator of hierarchical mergers than $\chieff$ --- a consequence of the generic properties of hierarchically-formed \acp{BBH}. 

The remainder of this letter is as follows. 
We outline the cluster population models used to construct the simulated set of first-generation (1G1G) and hierarchical \ac{BBH} mergers in Sec.~\ref{sec:cmc} before discussing how we quantify the measurements of the spin parameters in Sec.~\ref{subs:lr}. 
The results of this calculation using the simulated population of \ac{BBH} mergers as well as a selection of observed gravitational-wave signals are presented in Sec.~\ref{subs:res}. 
Finally, concluding remarks and implications of this study are presented in Sec.~\ref{sec:conc}. 

\section{Cluster population models}\label{sec:cmc}

We assemble our synthetic sample of dynamically-formed binary black hole mergers using the \texttt{CMC Cluster Catalog}, a suite of $N$-body cluster simulations spanning the parameter space of globular clusters observed in the local universe \citep{Kremer2020}. This catalog of models is computed using \texttt{CMC} \citep{Rodriguez2022}, a H\'{e}non-type Monte Carlo code that includes various physical processes specifically relevant to the dynamical formation of black hole binaries in dense star clusters including two-body relaxation, stellar and binary evolution \citep[computed using \texttt{COSMIC};][]{Breivik2020}, and direct integration of small-$N$ resonant encounters including post-Newtonian effects \citep{Rodriguez2018}. In total, this catalog contains 148 independent simulations with variation in total cluster mass, initial virial radius, metallicity, and cluster truncation due to galactic tidal forces. The chosen values for these parameters reflect the observed properties of the Milky Way globular clusters \citep[e.g.,][]{Harris1996}, but also serve as reasonable proxies for extragalactic clusters \citep[e.g.,][]{BrodieStrader2006} enabling a robust exploration of the formation of gravitational-wave sources in dense star clusters throughout the local universe.

To obtain a realistic astrophysically-weighted sample of binary black hole mergers, we follow \citet{RodriguezLoeb2018} and \citet{Zevin2021Eccentricity}: each of the 148 simulations are placed into equally-spaced bins in cluster mass and logarithmically-spaced bins in metallicity. Each cluster model is then assigned a relative astrophysical weight corresponding to the number of clusters expected to form in its associated 2D mass-metallicity bin across cosmic time, assuming that initial cluster masses follow a $\propto\,M^{-2}$ distribution \citep[e.g.,][]{LadaLada2003} and that metallicities (as well as corresponding cluster formation times) follow the hierarchical assembly distributions of \citet{ElBadry2019}. For all binary black hole mergers in a given model, the drawn cluster formation time is then added to the black hole binary's merger time, yielding a realistic distribution of binary black hole merger events as a function of redshift. This scheme yields a predicted binary black hole merger rate of roughly $20\,\rm{Gpc}^{-3}\,\rm{yr}^{-1}$ in the local universe from dense star clusters.

We account for detectability of the simulated binary systems by generating colored Gaussian noise corresponding to a three-detector LIGO-Virgo gravitational-wave detector network at both design sensitivity~\citep{Aasi2014, Acernese2015} and at the sensitivity the network achieved during the first half of \ac{LVK}'s third observing period ~\citep[O3;][]{GWTC2}. 
We then add the simulated signals, randomly generating the binary's orientation and sky position, to the detector network noise and calculate the matched-filter signal-to-noise ratio~\citep{Cutler1994}. 
Signals which pass the threshold signal-to-noise ratio (SNR) of ten are kept within the set of simulated detections. 

In the \texttt{CMC} simulations, all black holes formed via stellar evolution are assumed to have negligible birth spin, a reasonable assumption if angular momentum transport in their massive-star progenitors is highly efficient \citep[e.g.,][]{Qin2018, FullerMa2019}. However, spin can be imparted to cluster black holes through previous black hole merger events~\citep{Buonanno2008}. We assume all spin tilts are isotropically distributed. 
In addition to the non-spinning first-generation mergers, we consider two additional populations --- the population of hierarchical \acp{BBH} formed consistently from these non-spinning first-generation systems, and first-generation mergers with black hole spins artificially included between [0, 0.2]. 
The latter population is included as a ``worst-case'' scenario for first-generation mergers that are not formed with small spins. 
While we do not self-consistently generate a fourth population corresponding to hierarchical mergers from this spinning first-generation population, modifications to the spin properties of first-generation \acp{BH} only marginally change the distribution of hierarchical merger parameters (cf. Figs. 4, 6, and 7 from~\cite{Rodriguez2019}). 
The dominant impact of a spinning first-generation population is a significant reduction in the rate of hierarchical mergers, which does not affect our conclusions significantly regarding distinguishing the mergers within the hierarchical population but would affect their rates via the number of systems that are retained~\citep{Rodriguez2019, Mahapatra2021, Zevin2022retention}. 

\begin{figure}
    \centering
    \includegraphics{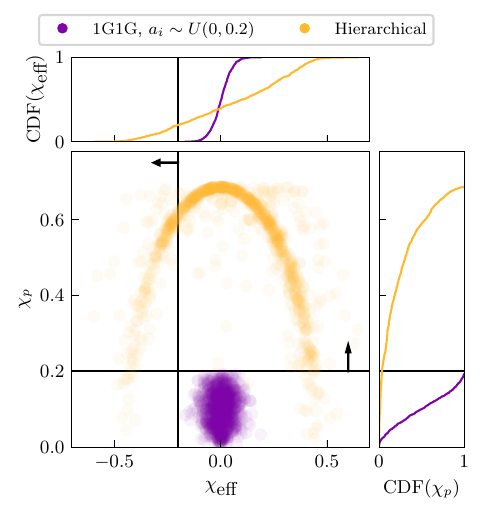}
    \caption{Two-dimensional distribution of spin parameters, $\chieff$ and $\chip$, for detectable low-spinning first-generation \acp{BBH} (1G1G; purple), and hierarchically formed \acp{BBH} (yellow).
    The one-dimensional marginal cumulative distribution functions (CDFs) are shown in the top and right panels.
    The spins of the low-spinning population are drawn uniformly and isotropically with spin-magnitudes from 0 to 0.2 in post-processing.
    All black hole masses are determined from the cluster simulations. 
    We have selected for signals that are detectable by enforcing a signal-to-noise ratio threshold of $10$ across the three detector LIGO-Virgo network at the \ac{LVK}'s sensitivity during their third observing period.
    The threshold of $\chit = 0.2$ used throughout the manuscript is indicated by the black lines for $\chieff$ and $\chip$. 
    A significantly greater fraction of the hierarchical systems possess $\chip > 0.2$ than $\chieff < -0.2$.}
    \label{fig:injections}
\end{figure}

In Fig.~\ref{fig:injections} we show the spin parameters, $\chieff$ and $\chip$, of the O3-detected set of simulations from the low-spinning first-generation (purple), and hierarchical \acp{BBH}.
The black lines indicate reasonable thresholds beyond which no 1G1G systems reside in the $\chieff$-$\chip$ parameter space. 
While $\chip$ is typically less well-measured in gravitational-wave observations~\citep{Ng2018, Biscoveanu2021}, hierarchical systems overwhelming produce more moderate-to-high $\chip$ \acp{BBH} and occupy a unique region of the $\chieff-\chip$ plane~\citep{Baibhav2021}. 
Therefore, in the following section, we explore the use of both $\chieff$ and $\chip$ as potential ``smoking-gun'' signatures of a \ac{BBH}'s hierarchical origin. 

\section{Distinguishing hierarchical mergers}

In this section, we turn our attention to how we might observationally identify the hierarchical mergers predicted from cluster populations using only the effective and precession spin parameters inferred from the observed GW signals. 
We first outline how we quantify the significance of the measurement before applying the calculation to the simulated cluster populations following Sec.~\ref{sec:cmc} in addition to a number of gravitational-wave events from the \ac{LVK}'s third observing period which may present evidence of hierarchical origin based on their leading-order spin measurements. 

\subsection{Quantifying spin measurement significance}~\label{subs:lr}

To understand the detectability of $\chip$ and $\chieff$ in the simulated populations produced in Sec.~\ref{sec:cmc}, we infer the 15 binary parameters (assuming quasi-circular orbits) for each merger injected into the two gravitational-wave networks considered. 
We then calculate the posterior distributions on $\chieff$ and $\chip$ directly from the inferred spin parameters, using Eqs.~\eqref{eq:chieff} and~\eqref{eq:chip}. 

To quantify how significantly $\chieff$ and $\chip$ are measured beyond the chosen thresholds, we utilize a ``likelihood-ratio''-based statistic, denoted $\textrm{LR}$. 
This threshold boundary is somewhat arbitrary but can be motivated from the cluster simulations in Sec.~\ref{sec:cmc}.
We compute $\textrm{LR}$ by integrating over the marginal single-event likelihood and a uniform prior bounded between the threshold and the parameter boundaries (here denoted $\chi_L$ and $\chi_U$ for the lower and upper edges respectively). 
For example, 
\begin{equation}\label{eq:LR}
    \textrm{LR}^{\chi > \chit}_{\chi \leq \chit} = \frac{\int_{\chit}^{\chi_U}\mathcal{L}(d|\chi)U(\chit, \chi_U)\,\dd\chi }{\int_{\chi_L}^{\chit} \mathcal{L}(d|\chi)U(\chi_L, \chit)\,\dd\chi}
\end{equation}
computes the likelihood-ratio for support above the threshold, $\chit$, compared to below the threshold. 
Here, $\mathcal{L}(d|\chi)$ is the marginal likelihood for the observed event data, $d$ given the spin parameter $\chi$ (either $\chip$ or $\chieff$). 
We use the analytical expressions from \cite{Callister2021a} to construct the marginal likelihood (i.e. all prior dependence, $\pi(\chi|q)$, is removed). 
It is important to note, however, that in marginalizing over all other degrees of freedom we have made implicit choices for the prior distributions on other parameters, such as the individual black hole masses and redshift. 
We use uniform-in-detector-frame component mass priors when sampling in chirp mass and mass ratio~\citep{Romero-Shaw2020, Callister2021a}, and a Euclidean luminosity distance prior ($\propto d_L^2$). 
While these choices will inevitably have an impact on the inferred LR values, we are aiming to identify \textit{unequivocally spinning} systems. 
Equation~\eqref{eq:LR} can also be inverted to compute the likelihood-ratio for support below the threshold. 

Upon close examination of Eq.~\eqref{eq:LR}, astute readers would note that it closely resembles a Bayes factor between two possible hypotheses (a spin parameter either above or below $\chi_{\rm thres}$)\footnote{We have opted for the terminology ``likelihood-ratio'' here as we are removing the explicit and complex behavior of the posterior distribution with respect to the prior. If interested, 
\begin{equation}
    \textrm{LR}^{\chi > \chit}_{\chi \leq \chit} = \frac{\int_{\chit}^{\chi_U}p(\chi|d)\,\dd\chi }{\int_{\chi_L}^{\chit} p(\chi|d)\,\dd\chi}
\end{equation}
could be computed instead, where $p(\chi|d)$ is the marginal posterior distribution.}. 
Therefore, we can interpret the inferred value in a similar way --- the likelihood-ratio quantifies the amount of support above (below) the threshold against the support below (above) it. 
A common metric in the field of Bayesian statistics is that a $\ln \textrm{LR}^{\chi > \chit}_{\chi \leq \chit} > 8$ quantifies significant evidence, corresponding to a $\sim3000\rm{:}1$ preference for $\chi > \chit$~\citep{Jefferys1961}. 
Due to the nature of this calculation, there is statistical uncertainty due to a finite number of posterior distribution samples above $\chit$. The uncertainty in $\ln\,\textrm{LR}$ scales approximately, ignoring the impact of the removal of the prior, as $\sim\sqrt{\textrm{LR}/N}$, where $N$ in the posterior samples. 
Since we have $\sim40,000$ samples per event, this corresponds to an uncertainty of $\sim0.3$ at $\ln\,\textrm{LR}=8$. This may slightly modify the exact percentage of systems passing the chosen $\ln\,\textrm{LR}=8$ threshold, though the broader conclusions of the Letter are unaffected.

There are, of course, many other parameters and methods to quantify this significance~\citep{Fairhurst2019b, Fairhurst2019a, Gerosa2020, Thomas2020}. 
Here we utilize this straight-forward approach for two reasons. 
The first is that it is intuitive to interpret from the one-dimensional marginal distribution -- \textit{how much support is above or below a threshold?}
And the second is that this statistic is more directly understood by the leading order terms in the gravitational-wave radiation due to both $\chieff$ and $\chip$, rather than being related first to the noise properties as in~\cite{Fairhurst2019b, Fairhurst2019a}.
Therefore, with a choice of spin threshold for the LR ($\chit$; motivated by Sec.~\ref{sec:cmc}) and under the assumption that all systems which pass $\chit$ are hierarchical mergers, we can use measurements of LR as a proxy for a definitive detection of a hierarchical merger. 
A $\chit$ value of 0.2 is motivated by confidently bounding observations above the expected small spins from \cite{Qin2018} and \cite{FullerMa2019}. We further choose more conservative thresholds ($\chit=0.3,0.4$) in the case where first-generation black holes might have some mechanism of being spun up~\citep[e.g.][]{Ma2023}. However, these systems still typically possess spins below $0.4$ and are rare~\citep[e.g. see App. A.1.3 of][]{Zevin2021OneChannel}. 
Additionally, it is expected that the presence of hierarchical mergers formed from first generation \ac{BBH} mergers with birth spins above 0.2 is heavily suppressed due to ejection of the merger remnant from the cluster environment~\citep{Rodriguez2019}. Finally, the more conservative bound of $\chit=0.4$ is consistent with the population observed thus far by the \ac{LVK}~\citep{GWTC3pop} being consistent with only first generation black holes.
This measure relies heavily on only the spins of the system, and so the statements in following sections are conservative. 
Information about the masses could be incorporated to boost the significance, though a threshold on masses will then need to be chosen as well, may be less motivated given large uncertainties in the underlying first-generation mass distributions~\citep[e.g.,][]{Mapelli2021,Spera2022}, and will inadvertently remove lighter hierarchical systems from consideration. 

\subsection{Application to cluster population models}~\label{subs:res}

\begin{figure*}
    \centering
    \includegraphics[width=\linewidth]{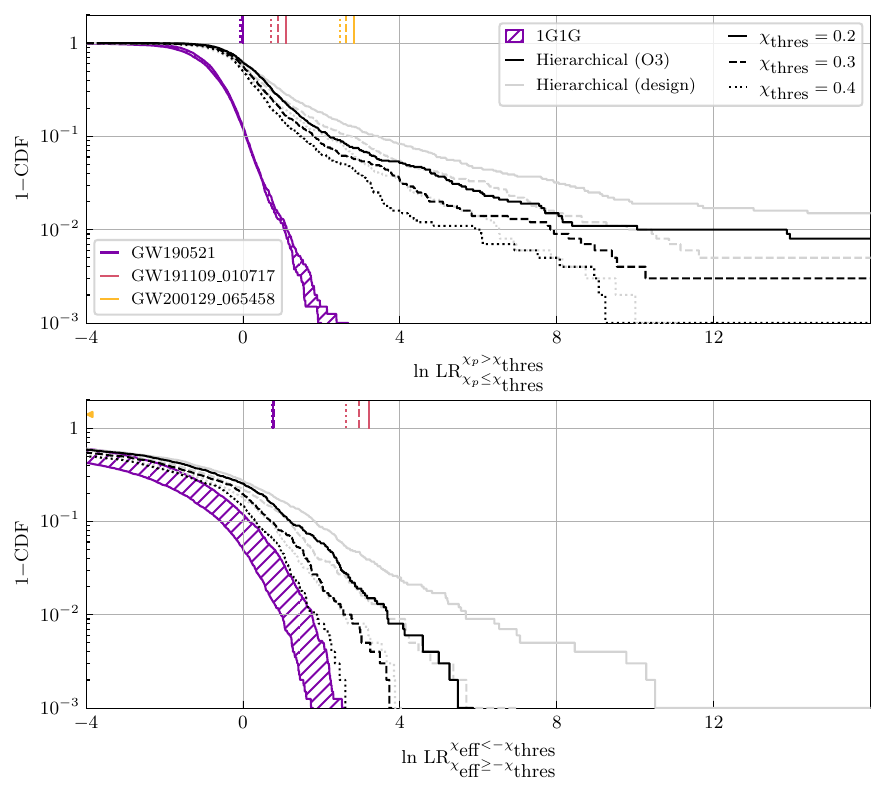}
    \caption{The complementary cumulative distribution function ($1-\textrm{CDF}$) of detectable 1G1G (shaded; purple) and hierarchical \ac{BBH} mergers (lines) as a function of the logarithmic likelihood ratio, $\ln\,\textrm{LR}$, defined in Eq.~\eqref{eq:LR}. 
    The three different linestyles correspond to different threshold choices ($\chit=0.2,0.3,0.4$), and shadings correspond to simulated signals detected in the first half of the \ac{LVK}'s third observing period (O3) sensitivity (dark), or a three-detector LIGO-Virgo network at design sensitivity (light). 
    The top and bottom panels correspond to the complementary cumulative distribution functions for $\chip$ and $\chieff$, respectively.
    Finally, the observed values of $\ln\,\textrm{LR}$ at the different thresholds for three gravitational-wave observations made during O3 --- GW190521 (purple), GW191109\_010717 (pink), and GW200129\_065458 (yellow) --- are marked. 
    A significantly larger fraction of the hierarchical population possess a confidently measurable value of $\chip$, whereas only the most relaxed threshold at design sensitivity can lead to a confident negative $\chieff$ measurement in a single event.}
    \label{fig:lnLR}
\end{figure*}

To explore how effectively hierarchical mergers can be selected out from a given population using spin parameters, we infer the properties of 1000 mergers from each of the three simulated populations (1G1G, 1G1G with uniform spin magnitudes in the range [0, 0.2], and hierarchical systems; as described in Sec.~\ref{sec:cmc}) in the current gravitational-wave detector network (from the third observing period; \citealt{GWTC2}) and at design sensitivity~\citep{Aasi2014, Acernese2015}. 
We simulate these signals using the gravitational waveform model {\tt \sc IMRPhenomXPHM}~\citep{Pratten2020}, which we add into Gaussian noise colored by the respective noise power spectral densities.
We arrive at 6000 posterior distributions\footnote{Publicly available posterior samples are available at \url{https://doi.org/10.5281/zenodo.10558308}.}, using the nested sampling algorithm \texttt{dynesty}~\citep{speagle2020dynesty} embedded within the Bayesian inference library \texttt{Bilby}~\citep{Ashton2018, Romero-Shaw2020}, from which we calculate the LR following Eq.~\eqref{eq:LR}. 
From these results, we can then construct the complementary cumulative distribution function indicating the recovered fraction of observations that have a LR above a given value. 
The result of this calculation is shown in Fig.~\ref{fig:lnLR}  for  both $\chip$ (top) and $\chieff$ (bottom). 
We find little difference in the inferred distribution of values of $\textrm{LR}$ for 1G1G systems, independent of detector sensitivity  and only slightly dependent on the choice of threshold and spin distribution. 
We therefore group all such possible distributions into the hatched purple region in Fig.~\ref{fig:lnLR}. 
The fraction of hierarchical binaries for different thresholds are shown in black and grey for \ac{LVK}'s gravitational-wave detector network at O3 sensitivity and at design sensitivity, respectively. 
The complementary cumulative distribution function as a function of the LR represents the fraction of simulated observations above a LR value. 
Finally, we also include the relevant values from three gravitational-wave observations with ticks above the curves: GW190521 (purple; \citealt{GW190521, GW190521Impl, GWTC2_1}), GW191109\_010717 (pink), and GW200129\_065458 (yellow; \citealt{GWTC3}). 

From Fig.~\ref{fig:lnLR}, we can identify the fraction of hierarchical binaries which pass a particular threshold of likelihood-ratio for both $\chip$ and $\chieff$.
Focusing on observations in the third \ac{LVK} observing period (O3), $\sim2\%$ of hierarchical mergers possess $\ln\,\textrm{LR}^{\chip > 0.2}_{\chip \leq 0.2} > 8$, indicating a confident detection.
Signal-to-noise ratio has a mild impact on the systems with high LRs, with higher SNR systems somewhat more likely to have a higher LRs. For example, $15\%$ of hierarchical systems have SNR $> 20$, whereas $63\%$ of all hierarchical systems with $\ln\,\textrm{LR}^{\chip > 0.2}_{\chip \leq 0.2} > 4$ possess an SNR $>20$. We anticipate much of the support for higher values of $\chip$ in these systems is also a product of clear imprints of spin precession in the waveform from specific spin configurations.
However, no choice of $\chit$ can provide a confident measurement for negative $\chieff$ except with the most liberal threshold ($\chit = 0.2$) at design sensitivity of the three-detector LIGO-Virgo detector network. 
Therefore, from the simulated population of binary black hole mergers from globular clusters, $\chieff$ is a wholly ineffectual parameter for distinguishing individual\footnote{This does not invalidate hierarchical studies where a population of potentially anti-aligned systems may be identified, as more information is extracted from a population of sources~\citep[e.g.][]{LVKpopsO3,Fishbach2022,Miller2024}.} hierarchical mergers\footnote{We also computed LR with the primary black-hole spin magnitude ($a_1$). We find that $\sim4\%$ of hierarchical mergers possess $\ln\,\textrm{LR}^{a_1 > 0.2}_{a_1 \leq 0.2} > 8$. While insightful, this does not factor in spin alignment and therefore such a measure may be contaminated by other channels.}.
Furthermore, if we instead treat the 1G1G population as a ``null'' background distribution from which to define a threshold (which is a very liberal threshold --- requiring complete confidence in the population model), we still arrive at similar conclusions. 
With a detection threshold informed from the 1G1G LR distribution ($\ln\,\textrm{LR}^{\chi > 0.2}_{\chi \leq 0.2} > 3$), we find $\sim8\%$ of hierarchical mergers would be distinguishable via precession effects, while only $\sim3\%$ would be distinguishable from $\chieff$ measurements. 
While we believe it to be difficult to claim any one observation is of a hierarchical origin with $\ln\,\textrm{LR}\sim3$, an ensemble of such observations would indicate some number of these observations were hierarchical. This may lead to hints at the level of a population of hierarchical \ac{BBH} mergers in the \ac{LVK}'s current fourth observing period --- even if we are not confident in the origin of any one event.

Finally, we briefly turn our attention to a select few events from the \ac{LVK}'s third observing period (O3) that have been discussed in the literature as potential systems with anti- or mis-aligned spins --- GW190521, GW191109\_010717, and GW200129\_065458~\citep{GW190521, GW190521Impl, GWTC2, GWTC2_1, GWTC3}.
For simplicity and direct comparison to the simulated mergers, we use only posteriors constructed using {\tt \sc IMRPhenomXPHM}\footnote{While GW190521~\citep{GWTC2} and GW200129\_065458~\citep{GWTC3, Hannam2021} have results with waveform models more closely resembling numerical relativity ({\tt \sc NRSur7dq4}; \citealt{Varma2019}), using these samples for these two results only marginally affects these conclusions.}. 
Using the LR calculation, no events surpass $\ln\,\textrm{LR} > 8$ for either $\chip$ or $\chieff$, although with a reduced threshold of $\ln\,\textrm{LR} > 3$, GW200129\_065458 and GW191109\_010717 pass the thresholds for $\chip$ and $\chieff$, respectively. 
However, since the impact of data quality issues impacting the interpretation of these events is still an open question, caution should be taken when interpreting these results~\citep[see][]{Davis2022, Payne2022, Udall2022, Macas2023}. 

\section{Conclusions}~\label{sec:conc}

Unequivocal detections of a hierarchical \ac{BBH} merger via gravitational-wave observations will help understand the formation channels and histories of such systems. 
While studies often focus on identifying a hierarchical merger from anti-aligned spins~\citep[see e.g.,][]{Zhang2023, Fishbach2022},
we have focused on both the measurement of spin-precession in addition to anti-alignment in a simulated \ac{BBH} merger population from realistic cluster models~\citep{Rodriguez2022}. 
From this study, the key insights are as follows:
\begin{enumerate}
    \item We have demonstrated that, in a realistic cluster population, \textit{determining a system to be hierarchical will likely first come from the measurement of spin-precession} (cf. Fig.~\ref{fig:lnLR}). 
    \item Additionally, from these simulated \ac{BBH} mergers from 1G1G and hierarchical systems, we can approximately discern the number of gravitational-wave observations needed to uncover a hierarchical system in such a manner. 
    We generally find that we should not yet have expected to confidently identify a hierarchical merger.
    Since $\sim25\%$ of the detectable \acp{BBH} from the cluster population are hierarchical, and $\sim2\%$ are confidently detectable at current sensitivity of the gravitational-wave network (from Fig.~\ref{fig:lnLR}), there is only a $25\%$ chance one or more hierarchical mergers would have been \textit{detectable} in the LVK's third observing run~\citep{GWTC2, GWTC3, GWTC2_1}. This probability should be considered a generous upper limit, as it assumes dynamical formation in globular clusters as the only channel and environment.
    \item Future observations appear much more fruitful. 
    At design sensitivity $\sim4\%$ of hierarchical mergers become distinguishable. 
    With an increased number of detections (ranging from $\sim$200--1000;~\citealt{Kiendrebeogo2023}), 
    one can reasonably expect $\sim2$--10 identifiably hierarchical systems. 
    Crucially, this analysis cannot be undertaken using anti-alignment of spins (i.e. $\chieff$), as such effects will not be detectable, even in the most optimistic of circumstances. 
\end{enumerate}

As the ground-based gravitational-wave detector network evolves and approaches its design sensitivity, the tangible possibility of observing a unequivocally spinning, hierarchical merger will become a reality. 
As we enter this era, the conclusions drawn here will be important in future discussions about the hierarchical origins of yet-to-be-detected \ac{BBH} mergers. 
When discussing such a system, in this Letter we find it will be significantly more advantageous to investigate the spin-precession than spin misalignment.
This motivates current and future research into both population modelling for hierarchical systems (and their first-generation progenitors) and waveform modeling to accurately capture this effect. 

\acknowledgments

The authors thank Zoheyr Doctor, Christopher Berry, and Parthapratim Mahapatra for insightful comments on the Letter. 
Support for KK was provided by NASA through the NASA Hubble Fellowship grant HST-HF2-51510 awarded by the Space Telescope Science Institute, which is operated by the Association of Universities for Research in Astronomy, Inc., for NASA, under contract NAS5-26555.
MZ gratefully acknowledges funding from the Brinson Foundation in support of astrophysics research at the Adler Planetarium.

The authors are grateful for computational resources provided by the LIGO Laboratory and supported by National Science Foundation Grants PHY-0757058 and PHY-0823459.
This material is based upon work supported by NSF's LIGO Laboratory which is a major facility fully funded by the National Science Foundation.
LIGO was constructed by the California Institute of Technology and Massachusetts Institute of Technology with funding from the National Science Foundation and operates under cooperative agreement PHY-1764464. This paper carries LIGO Document Number LIGO-P2400050.

\software{\texttt{Bilby}~\citep{Ashton2018, Romero-Shaw2020}; 
\texttt{dynesty}~\citep{speagle2020dynesty};
\texttt{iPython}~\citep{ipython}; 
\texttt{Matplotlib}~\citep{matplotlib}; 
\texttt{NumPy}~\citep{numpy}; 
\texttt{Pandas}~\citep{pandas};
\texttt{SciPy}~\citep{scipy}.}


\bibliography{library}{}
\bibliographystyle{aasjournal}

\end{document}